\documentclass[12pt,preprint]{aastex}

\newcommand{\be}{\begin{equation}}
\newcommand{\ee}{\end{equation}}

\newcommand{\bea}{\begin{eqnarray}}
\newcommand{\eea}{\end{eqnarray}}

\newcommand{\um}{\hbox{$\mu$m}}

\newcommand{\nWmmsr}{\hbox{nW m$^{-2}$ sr$^{-1}$}}

\newcommand{\etal}{{\it et al.\/}}
\newcommand{\vs}{{\it vs.\/}}

\begin{document}

\title{Comparing Optical and Near Infrared Luminosity Functions}

\author{Edward L. Wright\altaffilmark{1}} \affil{Institute for Advanced Study,
Princeton NJ 08540}
\altaffiltext{1}{%
Permanent address: Department of Physics and
Astronomy, University of California, Los Angeles, CA  90095-1562}
\email{wright@astro.ucla.edu}

\begin{abstract}
The Sloan Digital Sky Survey [SDSS] has measured an optical luminosity
function for galaxies in 5 bands, finding 1.5 to 2.1 times more luminosity
density than previous work.  This note compares the SDSS luminosity
density to two recent determinations of the near infrared luminosity
function based on 2MASS data, and finds that an extrapolation of the SDSS
results gives a 2.3 times greater near infrared luminosity density.
\end{abstract}


\section{Introduction}

The current luminosity density of the Universe is an important input
into estimates of both the chemical evolution of the Universe due to nuclear 
burning in stars, and the extragalactic background light or cosmic
optical and infrared background.  The luminosity density is found by
an integral over the luminosity function of galaxies, and
estimating the luminosity function of galaxies requires both
photometric measurements of galaxies and redshifts for distances.  The
recent release by 2MASS \citep{2MASS-Exp-Sup} of near infrared
photometric data covering one-half the sky has led to two recent
estimates of the near IR luminosity function: \citet{KPFHM01} measured
redshifts or used existing redshifts for a bright sample over a large
[$\sim 2$ sr] region on the sky, while \citet{2dF2MASSLF} analyze a
deeper sample over an effective area of $\sim 0.2$ sr using redshifts
from the 2 Degree Field Galaxy Redshift Survey [2dFGRS].  The massive
optical photometric and spectroscopic Sloan Digital Sky Survey [SDSS]
has also estimated the luminosity function of galaxies in five bands
between 0.3 and 1 \um\ \citep{SDSSLF}.  The SDSS luminosity density is
1.5 to 2.1 times higher than previous estimates of the optical
luminosity density.  \citet{SDSSLF} go through the exercise of
evaluating galaxy fluxes in the ways identical to the methods
previously used, and found that the larger flux seen by the SDSS was
due to the larger apertures employed by the SDSS.

This paper extrapolates this luminosity density comparison of the SDSS
to other determinations into the near infrared.

\section{Luminosity Density}

\begin{figure}[tbp]
\plotone{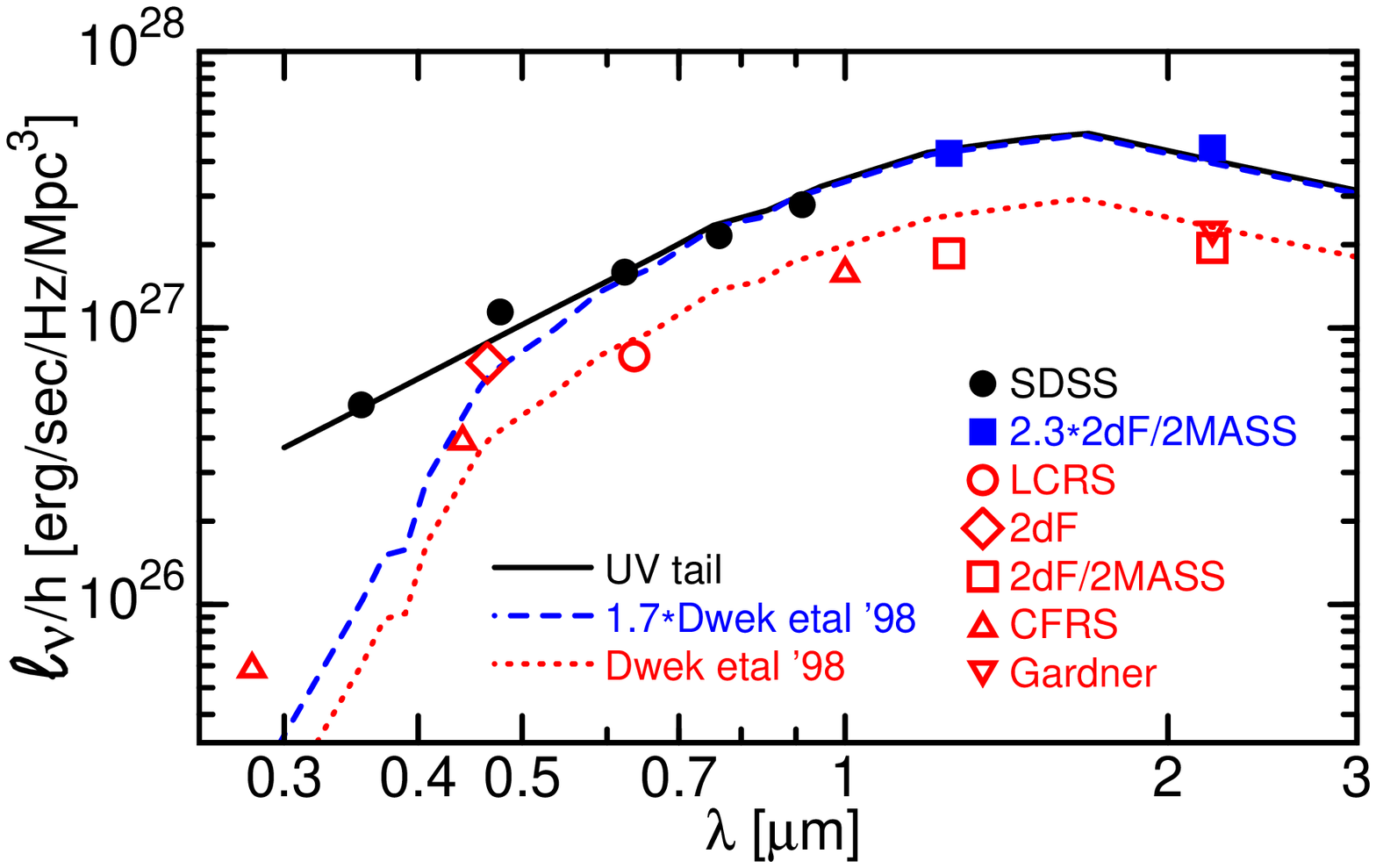}
\caption{%
Luminosity density per unit frequency
$\ell_\nu$ computed from the SDSS \citep{SDSSLF},
LCRS \protect\citep{1996ApJ...464...60L},
2dF \protect\citep{1999MNRAS.308..459F}, and
2dF/2MASS \protect\citep{2dF2MASSLF}
CFRS \protect\citep{1996ApJ...460L...1L} and
\protect\citet{1997ApJ...480L..99G} luminosity functions.
Dividing by $h = H_\circ/100$ makes this quantity independent of the
Hubble constant.
The dashed curve is the mean spiral spectrum
from Figure 5 of \protect\citet{1998ApJ...508..106D}, 
but scaled by a factor of 1.7, while the solid curve is the same function
with an $\ell_\lambda=$~const [or $\ell_\nu \propto \nu^{-2}$] 
tail starting at 0.8~\um.
The dotted curve shows this function with its original normalization.
\label{fig:lumden}}
\end{figure}

\begin{table}[tbhp]
\begin{center}
\begin{tabular}{lrrrrrrr}
\hline
$\lambda\;[\um]$     & 0.354 & 0.477 & 0.623 & 0.763 & 0.913 & 1.25 & 2.20 \\
$\phi_*$ [Mpc$^{-3}$] & 0.0400 & 0.0206 & 0.0146 & 0.0128 & 0.0127 & 0.0104 &
0.0108 \\
$\alpha$ & -1.35 & -1.26 & -1.20 & -1.25 & -1.24 & -0.93 & -0.96 \\
$M_*$ & -18.34 & -20.04 & -20.83 & -21.26 & -21.55 & -21.40 & -21.57 \\
\hline
\end{tabular}
\end{center}
\caption{Schechter luminosity function parameters used in the Figures,
taken from \protect\citet{SDSSLF} and \protect\citet{2dF2MASSLF}.
$M_*$, the absolute magnitude at $L_*$, is given on the AB system.
\label{tab:LFpar}}
\end{table}

While luminosity functions can be defined in many ways, using either
parametric models or non-parametric estimators, this paper only compares the
\pagebreak[1]
parametric \citet{SchechterLF} luminosity function fits.
The Schechter luminosity function is given by
\be
n(L)dL = \phi_* (L/L_*)^\alpha \exp(-L/L_*) dL/L_*
\ee
and it gives a luminosity density
$\ell = 4\pi j = \int L n(L) dL = \phi_* L_* \Gamma(\alpha+2)$.
This luminosity density is plotted as function of wavelength for the
five SDSS bands 
from \citet{SDSSLF} and the near infrared J and K bands 
from \citet{2dF2MASSLF} in
Figure \ref{fig:lumden} along with results from earlier optical and near
infrared studies.
Table \ref{tab:LFpar} lists the parameters of the Schechter fits used to
compute the luminosity density.
Note that the measured luminosity densities scale like the Hubble constant
$H_\circ$, so the plotted densities have been divided
by a factor of $h = H_\circ/100$.
It is clear that there is a large discontinuity between the optical SDSS
bands and the near infrared bands.
Also shown on Figure \ref{fig:lumden} is the mean spiral galaxy spectrum
from \citet{1998ApJ...508..106D}.
\pagebreak[1]
A least sum of absolute errors fit to a model that allowed different
scaling factors in the optical and near infrared, and also allowed for
an $\ell_\lambda = \mbox{const}$ tail to the spectrum at short wavelengths,
gave the heavy curve in Figure \ref{fig:lumden}.  The relative scaling
between the SDSS and the 2MASS results was a factor of 2.3.
The overall scaling was a factor of $1.7 \times$ increase to get the
\citet{1998ApJ...508..106D} luminosity density to match the level of
the SDSS data.

\begin{figure}[tb]
\plotone{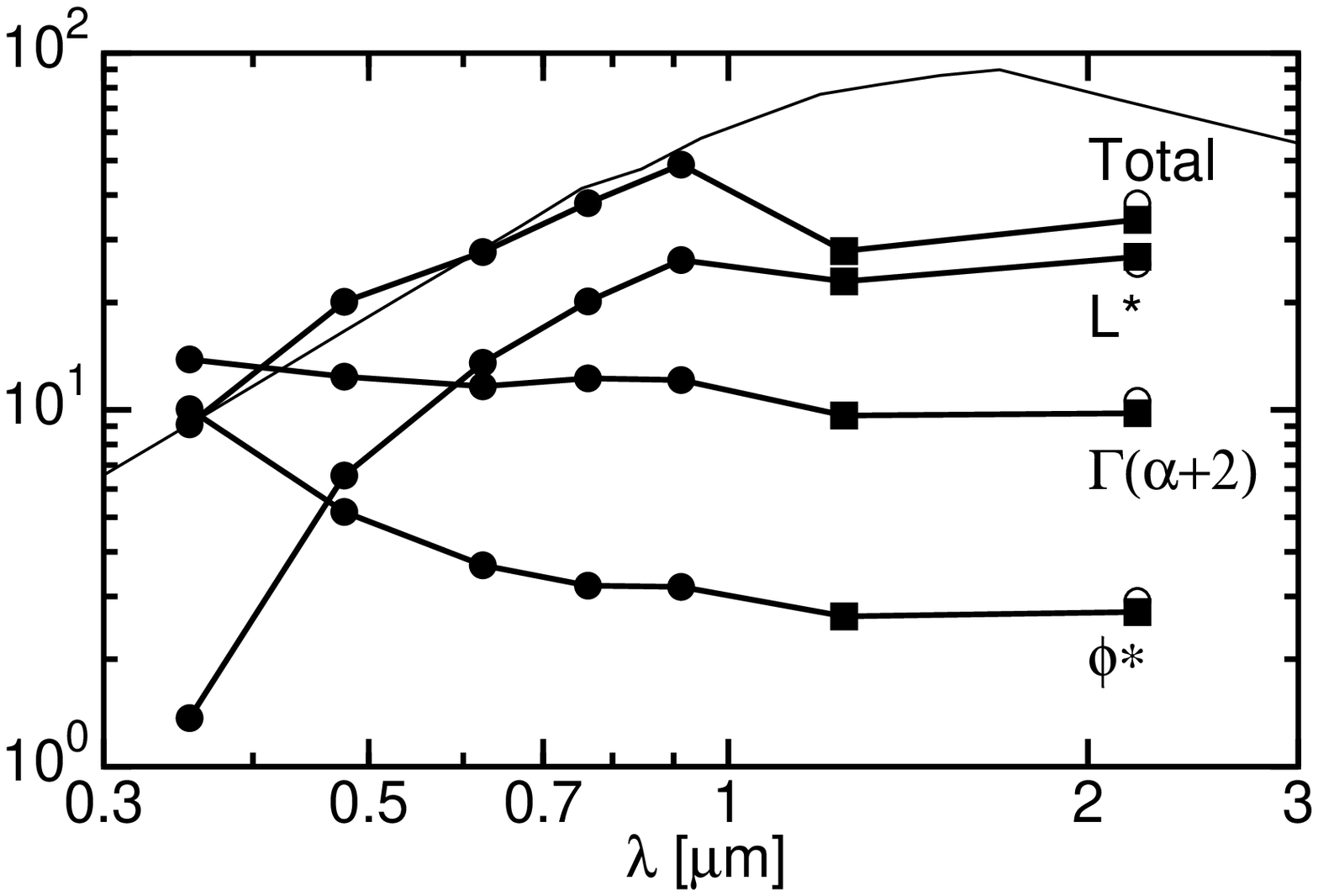}
\caption{%
Plot of the three factors in the luminosity density for the SDSS
(filled circles), the 2dF/2MASS (filled squares) and the
\protect\citet{KPFHM01} (open circle) luminosity functions.
The units on the $y$-axis are arbitrary, and each factor has been
scaled to fit on the plot.
The thin solid curve is the fit from Figure \protect\ref{fig:lumden}.
\label{fig:LFparams}}
\end{figure}

In order to study the source of the $2.3 \times$ discrepancy between
the SDSS and the 2MASS luminosity function,  Figure \ref{fig:LFparams}
plots both the luminosity density and the three factors that go into
the luminosity density as a function of wavelength.  This plot clearly shows
that the majority of the effect is caused by differences in $L_*$.
This plot also shows the \citet{KPFHM01} 
``all galaxy'' fit at 2.2 \um, which is quite consistent
with the \citet{2dF2MASSLF} fits.

\section{Discussion}

The local luminosity density difference of a factor of 1.5 between the SDSS
and the 2dF \citep{1999MNRAS.308..459F},
and a factor of 2.1 between the SDSS and the LCRS \citep{1996ApJ...464...60L},
extends into the near infrared where an extrapolated SDSS luminosity density
is a factor of 2.3 higher than recent determinations using 
2MASS \citep{2dF2MASSLF,KPFHM01}.  
The cause of the discrepancy between the SDSS and the LCRS and 2dF optical
luminosity density determinations is the use of larger apertures by the
SDSS \citep{SDSSLF}, but the cause of the difference in the near infrared
has not been determined.
If in fact
the optical and near infrared luminosity density has been underestimated
by a factor of about 2, this provides a partial explanation for the
high extragalactic background levels seen by
\citet{1999hrug.conf..487B} in the optical; and
\citet{2000ApJ...536..550G}, and \citet{2000ApJ...545...43W} in the near
infrared.
The background is given by
\be
4\pi J_\nu = \int \ell_{(1+z)\nu}(z) \; \frac{cdt}{dz} \; dz,
\ee
and while the luminosity functions discussed here only determine $\ell$ at
$z \approx 0$, one expects that an increase in $\ell$ at low redshift
would propagate smoothly to higher redshifts as well leading to an
increased background.

The existing number counts $N(>S)$, where $N$ is the number
of sources per steradian brighter than flux $S$, do not give a background
$J = \int S dN$
as large as the observed backgrounds: $\nu J_\nu = 8\pm2\;\nWmmsr$
at $2.2\;\um$ from counts \citep{2000MNRAS.312L...9M}
\vs\ $20\pm6\;\nWmmsr$ from DIRBE \citep{Wr01}.
The number counts for bright sources, in the Euclidean regime with
$N(>S) \propto S^{-3/2}$, are related to the local luminosity function by
\be
N(>S) = \frac{1}{3} (4\pi)^{-3/2} S^{-3/2} \int L^{3/2} n(L) dL
= \frac{\phi_* \Gamma(\alpha+2.5)}{3} \left(\frac{L_*}{4\pi}\right)^{3/2}
S^{-3/2}.
\ee
The bright end of the \citet{2000MNRAS.312L...9M} number counts at $2.2\;\um$
do agree with the \citet{2dF2MASSLF} determination of the luminosity function.
Increasing the flux of each galaxy by using the larger SDSS apertures
would increase $S$ at constant $N$ and thus explain the observed 
optical and near infrared backgrounds.
The integral under the solid curve in Figure \ref{fig:lumden}
gives a NIR-optical luminosity density of
$5.6h \times 10^8\;L_\odot/\mbox{Mpc}^3$.
This value is based on the SDSS data and the extrapolation of the SDSS data
into the near infrared shown in Figure \ref{fig:lumden}.

In a private E-mail, Shaun Cole suggested that the optical to near infrared
colors of galaxies in common between the SDSS and 2MASS samples are normal
and agree with the shape of the \citet{1998ApJ...508..106D} curve
in Figure \ref{fig:lumden}; and that he did not expect that the 2MASS
magnitudes missed a significant fraction of the flux - certainly not a
factor of 2.3.  If the SDSS luminosity densities are correct,
this would suggest that a significant fraction of the galaxies in the SDSS
catalog that should have been detectable in the near infrared
were missed by 2MASS.

On the other hand, the SDSS luminosity densities could be too high.  
If so, the
optical and near infrared background determinations could be too high,
or an exotic source such as a decaying elementary particle 
[see the ``DP'' curve in Figure 1b of \citet{1986ApJ...306..428B}]
could provide part of the background.
If the background determinations are too high, the
most likely cause of the error would be the zodiacal light modeling.
In the near IR, replacing the ``very strong no-zodi'' model
\citep{1997AAS...191.8703W,1998ApJ...496....1W,2000ApJ...536..550G}
with the \citet{1998ApJ...508...44K} model would {\em increase} the
background, making the discrepancy larger.
For example, \citet{CRCJ01} use the \citet{1998ApJ...508...44K} model and
get $\nu J_\nu = 28\pm7\;\nWmmsr$ at $2.2\;\um$.

At least some -- if not all -- of the luminosity function and background
determinations discussed in this note
are incorrect.  In order to find out the true luminosity density from 
galaxies one will have to measure the total flux from galaxies in large 
apertures that include the low surface brightness fuzzy fringes.

\acknowledgments

Astrophysics research at the IAS is supported by National Science
Foundation Grant PHY-0070928 and the Ambrose Monell Foundation.
I would like to thank John Peacock, Shaun Cole, Michael Strauss and
Zeljko Ivezic for useful discussions on this paper.

\end{document}